\documentclass{ws-procs9x6}

\usepackage{amsbsy}
\makeatletter
\def\lsim{\mathrel{\mathpalette\@versim<}}
\def\gsim{\mathrel{\mathpalette\@versim>}}
\def\@versim#1#2{\vcenter{\offinterlineskip
        \ialign{$\m@th#1\hfil##\hfil$\crcr#2\crcr\sim\crcr } }}
\makeatother

\def\cal#1{\mathcal{#1}}

\def\pl{P}
\def\R{{\rm R}}
\def\L{{\rm L}}
\def\pl{{\rm Pl}}
\def\m1{{-1}}

\def\VEV#1{\left\langle#1\right\rangle}

\newcommand{\nn}{\nonumber\\}

\def\gam#1{\,{}^{#1}\kern-1pt\Gamma}

\def\ppmatrix#1{\left(\begin{array}{@{\,}c@{\,\ }c@{\,\ }c@{\,}}
#1\end{array}\right)}
\def\tpmatrix#1{\left(\begin{array}{@{\,}c@{\,\ }c@{\,}}
#1\end{array}\right)}

\begin{document}

\title{Impact and Implication of Bi-Large Neutrino Mixings on GUTs\footnote{
\uppercase{T}alk presented at \uppercase{NOON}2003 held at 
\uppercase{K}anazawa, \uppercase{F}eb.~10 -14, 2003.
}} 

\author{Taichiro KUGO\footnote{
\uppercase{W}ork partially supported by 
\uppercase{G}rant-in-\uppercase{A}id for \uppercase{S}cientific 
\uppercase{R}esearch \uppercase{N}o.~13640279 
from \uppercase{J}apan \uppercase{S}ociety for the 
\uppercase{P}romotion of \uppercase{S}cience, and 
\uppercase{G}rants-in-\uppercase{A}id for \uppercase{S}cientific 
\uppercase{R}esearch on \uppercase{P}riority \uppercase{A}rea  
``\uppercase{N}eutrinos" (\uppercase{Y}.~\uppercase{S}uzuki) 
\uppercase{N}o.~12047214 
from the \uppercase{M}inistry of \uppercase{E}ducation, 
\uppercase{S}cience, \uppercase{S}ports and \uppercase{C}ulture, 
\uppercase{J}apan.}}

\address{Yukawa Institute for Theoretical Physics, 
Kyoto University, \\ Kyoto 606-8502, Japan \\
E-mail: kugo@yukawa.kyoto-u.ac.jp}


\maketitle

\abstracts{
Under the assumptions that 1) the quark/lepton mass matrices take 
Froggatt-Nielsen's factorized power form $\lambda^{\psi_i+\psi_j}$ 
with anomalous $U(1)$ charges $\psi_i$, and 2) the $U(1)$ charges $\psi_i$ 
respect the $SU(5)$ GUT structure, we show that 
the quark mass data necessarily implies the large 2-3 mixing in the 
MNS mixing matrix $U_{\rm MNS}$. 
If we further add the data of the mass squared difference ratio of 
solar and atmospheric neutrinos, then, it implies that 
the 1-2 mixing in $U_{\rm MNS}$ is also large, 
so explaining the bi-large mixing. This analysis also 
gives a prediction that 
$U_{e3}\equiv(U_{\rm MNS})_{13}$ should be of order $\lambda\,\sim\,
(0.1 - 0.5)$.}

\section{Introduction}

Existence of a certain grand unified theory (GUT) beyond the standard 
model is guaranteed by i) the anomaly cancellation between quarks and 
leptons and ii) the unification of the gauge coupling constants at 
energy scale around $\mu\,\sim\,10^{15-16}$\,GeV. The strongest candidate 
for the unified gauge group is $E_6$, which is not only 
suggested by string theory but also unique in the property that it is 
the maximal safe simple group allowing 
complex representations in the $E$-series; $E_3=SU(3)\times SU(2)$, 
$E_4=SU(5)$, $E_5=SO(10)$, $E_6, \ E_7,\ E_8$.\cite{ref:Georgi}

The purpose of this talk is to analyze the implications of the neutrino 
data on the possible GUTs. 
This is based on a work\cite{BK:bi-large} in collaboration with 
Masako Bando.
The particular facts of the neutrino data 
are:\cite{SuperKam:1998,SuperKam:2002,SNO,Kamland}
\begin{enumerate}
\item Bi-large mixing
\begin{equation}
\sin^22\theta_{12} \  \sim\ (0.86 - 1.0), \qquad 
\sin^22\theta_{23} \  \sim\ 1.
\end{equation}
\item Mass-squared difference ratio
\begin{equation}
{\triangle m_\odot^2\over\triangle m_{\rm atm}^2} 
\ \sim\ {7\times10^{-5}\,{\rm eV}^2\over 
3\times10^{-3}\,{\rm eV}^2} 
\ \sim\ \lambda^{2 \hbox{-} 3}.
\label{eq:dmRatio}
\end{equation}
where $\lambda$ defined below is a quantity of magnitude 
$\lambda\sim 0.22$.
\end{enumerate}
These show a sharp contrast to the quark sector, in which 
the mixings are very small and the mass spectrum is hierarchical. 
The mutual relations of masses and mixing angles between 
quarks and leptons/neutrinos will be great clues for the GUTs. 

As an working hypothesis we here assume an supersymmetric SU(5) GUT and 
the Froggatt-Nielsen mechanism\cite{Froggatt:1978nt} 
to generate effective Yukawa coupling matrices of the form 
\begin{equation}
y\Psi_i\Psi_jH\left({\Theta\over M_\pl}\right)^{\psi_i+\psi_j+h},
\label{eq:FNcoupling}
\end{equation}
where the Yukawa couplings $y$ can in principle depend on the generation 
label $i,j$ but are assumed to be all order 1 and so are denoted by $y$ 
collectively. $\Theta$ is the Froggatt-Nielsen field carrying the $U(1)_X$ charge $-1$ 
and the $U(1)_X$ charges of the other Higgs field $H$ and matter fields 
$\Psi_i$ ($i=1,2,3$) are denoted by the corresponding lower-case letters: 
\begin{equation}
X(\Theta)=-1, \qquad X(H)=h, \qquad X(\Psi_i)=\psi_i\,(\,\geq0).
\end{equation}
After the  Froggatt-Nielsen field $\Theta$ develops a vacuum expectation 
value (VEV) $\VEV{\Theta}$, which is assumed to be smaller than 
the Planck scale by a factor of Cabibbo angle $\theta_{\rm C}$
\begin{equation}
{\VEV{\Theta}\over M_\pl} \equiv \lambda \ \sim \ 0.22\simeq \sin\theta_{\rm C},
\end{equation}
the effective Yukawa couplings induced from Eq.~(\ref{eq:FNcoupling})
are given by
\begin{equation}
y_{ij}^{\rm eff}= y \times \lambda^{\psi_i+\psi_j+h}= {\cal O}(1)\times\lambda^{\psi_i+\psi_j+h}.
\end{equation}
That is, the mass matrix $M$ takes the form
\begin{equation}
M = yv\lambda^h \times \bordermatrix{
 & &  \stackrel{j}{\vee} & \cr
 & &  & \cr
i{>} & & \lambda^{\psi^\R_i+\psi^\L_j}  & \cr
 & &  & \cr}
\end{equation}
with $\VEV{H}=v$. $\psi^\R_i$ and $\psi^\L_j$ are the $U(1)_X$ charges 
of the right-handed and left-handed matter fields $\Psi_i^\R$ and 
$\Psi_i^\L$, respectively.
Thus, in this Froggatt-Nielsen mechanism, the hierarchical mass structure can be 
explained by the difference of the $U(1)_X$ charges $\psi^{\R,\L}_i$ 
of the matter fields. Note that this type of `factorized' mass matrix 
can be diagonalized as 
\begin{equation}
VMU^\dagger = M^{\rm diag.}
\end{equation} 
by unitary matrices $V$ and $U$ taking also a similar power forms:
\begin{equation}
U \ \sim \ 
\bordermatrix{
 & & \stackrel{j}{\vee} & \cr
 & &  & \cr
i{>} & & \lambda^{\left|\psi^\L_i-\psi^\L_j\right|}  & \cr
 & &  & \cr}\ , \qquad 
V \ \sim \ 
\bordermatrix{
 & & \stackrel{j}{\vee} & \cr
 & &  & \cr
i{>} & & \lambda^{\left|\psi^\R_i-\psi^\R_j\right|}  & \cr
 & &  & \cr}
\end{equation}

\section{$U(1)_X$ charge assignment}

I assume $SU(5)$ structure at least for the $U(1)_X$ charge assignment. 
Then, first, we consider the Yukawa coupling 
responsible for the up-quark sector masses.
In order for the effective Yukawa coupling
\begin{eqnarray}
&&y_u\Psi_i({\bf10})\Psi_j({\bf10})H_u({\bf5})
\left({\Theta\over M_\pl}\right)^{\psi_i({\bf10})+\psi_j({\bf10})+h_u} 
\nn && \hspace{2em} 
\rightarrow \ 
{y_u}_{ij}^{\rm eff}=y_u \times \lambda^{\psi_i({\bf10})+\psi_j({\bf10})+h_u} 
\end{eqnarray}
to reproduce the observed up-type quark mass hierarchy structure 
\begin{equation}
m_t : m_c : m_u = 1 : \lambda^4 : \lambda^7\ ,
\end{equation}
we are led to choose the following values for the $U(1)_X$ charges of 
three generation $\Psi_i({\bf10})$ fermions 
taking $h_u=0$ for simplicity:\cite{ref:BK}
\begin{equation}
(\,\psi_1({\bf10}),\ \psi_2({\bf10}),\ \psi_3({\bf10})\,)
=(\,3,\ 2,\ 0\,)
\label{eq:value10}
\end{equation}

Next we consider the mass matrices of down-type quarks and charged 
leptons which come from the couplings
\begin{eqnarray}
&&y_d\Psi_i({\bf10})\Psi_j({\bf5}^*)H_d({\bf5}^*)
\left({\Theta\over M_\pl}\right)^{\psi_i({\bf10})+\psi_j({\bf5}^*)+h_d}
\nn && \hspace{2em} 
 \rightarrow \ 
{y_d}_{ij}^{\rm eff}= y_d \times \lambda^{\psi_i({\bf10})+\psi_j({\bf5}^*)+h_d}. 
\end{eqnarray}
Note that this yields the transposed relation between the 
 down-type quark mass matrix $M_d$ and the charged lepton one $M_l$:
$
{M_d}^T \ \sim\  M_l 
$. 
This is because the $\Psi_i({\bf5}^*)$ multiplets contain the 
right-handed component $d^c$ for the down-type quarks while 
the left-handed component $l$ for the charged leptons.
Therefore the unitary matrices for diagonalizing those mass matrices, 
satisfy the relations
\begin{equation}
\left\{
\begin{array}{@{\,}l}
V_dM_dU_d^\dagger = M_d^{\rm diag.} \\[.5ex]
V_lM_lU_l^\dagger = M_l^{\rm diag.}
\end{array}\right. \quad \rightarrow\quad 
\left\{
\begin{array}{@{\,}l}
V_l=U_d^* \\[.5ex]
V_d=U_l^* 
\end{array}\right. ,
\end{equation}
so that we have
\begin{equation}
U_d^* (M_l\sim M_d^T) U_l^\dagger = \hbox{diag.}\qquad \hbox{with}\quad 
\left\{
\begin{array}{@{\,}l}
U_d \ \sim\ \bigl( 
\lambda^{\left|\psi_i({\bf10})-\psi_j({\bf10})\right|}\bigr) \\[.5ex]
U_l \ \sim\ \bigl( 
\lambda^{\left|\psi_i({\bf5}^*)-\psi_j({\bf5}^*)\right|}\bigr) 
\end{array}\right. \,.
\end{equation}
That is, the mass matrix takes the form
\begin{equation}
M_d^T \sim M_l \  \sim \ 
yv\lambda^{h_d} \times \bordermatrix{
 & {\bf5}^*_1 & {\bf5}^*_2 & {\bf5}^*_3 \cr
{\bf10}_1 & \lambda^{3+\psi_1({\bf5}^*)} & \lambda^{3+\psi_2({\bf5}^*)} & \lambda^{3+\psi_3({\bf5}^*)}\cr
{\bf10}_2 &\lambda^{2+\psi_1({\bf5}^*)} & \lambda^{2+\psi_2({\bf5}^*)} & \lambda^{2+\psi_3({\bf5}^*)}\cr
{\bf10}_3 &\lambda^{\psi_1({\bf5}^*)} & \lambda^{\psi_2({\bf5}^*)} & \lambda^{\psi_3({\bf5}^*)}\cr }.
\label{eq:dlmass}
\end{equation}
In order for this $M_d$ to reproduce the mass ratio of the top and bottom 
quarks
\begin{equation}
{m_b\over m_t} \ 
 \mathrel{\mathop{\kern0pt \sim}\limits_{\hbox{\scriptsize exp.}}}\ 
\lambda^{2 - 3}
\end{equation}
we take $\psi_3({\bf5}^*)=2-h_d$.
Further, to reproduce 
the down-type quark mass hierarchy
\begin{equation}
m_b : m_s : m_d 
 \mathrel{\mathop{\kern0pt =}\limits_{\hbox{\scriptsize exp.}}}\ 
= 1 : \lambda^2 : \lambda^4\ ,
\end{equation}
we take $\psi_2({\bf5}^*)=2-h_d$ and $\psi_1({\bf5}^*)=3-h_d$; 
thus, we have
\begin{equation}
(\,\psi_1({\bf5}^*),\ \psi_2({\bf5}^*),\ \psi_3({\bf5}^*)\,)
=(\,3-h_d,\ 2-h_d,\ 2-h_d\,), 
\label{eq:value5}
\end{equation}
and the mass matrix (\ref{eq:dlmass}) now reduces to 
\begin{equation}
M_d^T \sim M_l \  \sim \ 
yv\lambda^2 \times 
\ppmatrix{
\lambda^4 & \lambda^{3} & \lambda^{3} \cr
\lambda^{3} & \lambda^{2} & \lambda^{2} \cr
\lambda    &  1  & 1 \cr} 
\end{equation}
This form of mass matrix is called lopsided.

\section{Mixing matrices}

Mixing matrices in the quark sector and lepton sector are called 
Cabibbo-Kobayashi-Maskawa (CKM) and Maki-Nakagawa-Sakata (MNS)\cite{ref:MNS} 
matrices and they are given by 
\begin{equation}
U_{\rm CKM} = U_uU_d^\dagger\,,\qquad 
U_{\rm MNS} = U_lU_\nu^\dagger\,.
\end{equation} 
In our case both $U_u$ and $U_d$ takes the form 
$U_u\,\sim\,U_d\,\sim\,
( \lambda^{\left|\psi_i({\bf10})-\psi_j({\bf10})\right|})$, 
so that the CKM matrix, generally, also has the same form
\begin{equation}
U_{\rm CKM} = U_uU_d^\dagger\ \sim\ 
\bigl( \lambda^{\left|\psi_i({\bf10})-\psi_j({\bf10})\right|}\bigr)
\ \sim\ 
\ppmatrix{
1 & \lambda & \lambda^3 \\
\lambda & 1 & \lambda^2 \\
\lambda^3 & \lambda^2 & 1 }.
\end{equation}

This is all right.  For the charged lepton sector we have 
\begin{equation}
U_l \ \sim\ 
\bigl( \lambda^{\left|\psi_i({\bf5}^*)-\psi_j({\bf5}^*)\right|}\bigr)
\ \sim\ 
\ppmatrix{
1 & \lambda & \lambda \\
\lambda & 1 & 1  \\
\lambda & 1 & 1 }.
\end{equation}
If the mixing matrix $U_\nu$ in neutrino sector is $\sim1$, 
this beautifully explains the observed 
large 2-3 neutrino mixing!  However, this alone fails in explaining 
the large 1-2 mixing. We thus have to discuss the 
neutrino mixing matrix $U_\nu$ now. 

%

\section{Neutrino mass matrix and mixing}

Generally in GUTs, there appear some right-handed neutrinos 
$\Psi_I({\bf1})=\nu_{\R I}$ ($I=1,\cdots,n$); 
for instance, $n=3$ in $SO(10)$ and $n=6$ in $E_6$. They will 
generally get superheavy Majorana masses denoted by an $n\times n$ 
mass matrix $(M_\R)_{IJ}$, 
and also possesses the Dirac masses (R-L transition mass terms) 
\begin{equation}
\bigl(M_{\rm D}^T\bigr)_{iI} \  \sim \ 
y_\nu v\lambda^{h_u} \times \bigl(
\lambda^{\psi_i({\bf5}^*)+\psi_I^\R} \bigr)
\end{equation}
induced from 
\begin{equation}
y_\nu\Psi_i({\bf5}^*)\Psi_I({\bf1})H_u({\bf5})
\left({\Theta\over M_\pl}\right)^{\psi_i({\bf5}^*)+\psi_I^\R+h_u}
\hspace{-2ex} \rightarrow \ 
{y_\nu}_{ij}^{\rm eff}= y_\nu \times \lambda^{\psi_i({\bf5}^*)+\psi_I^\R+h_u} 
\hspace{-.5ex}.\, 
\end{equation}
Here $\psi_I^\R$ denotes the $U(1)_X$ charges of the right-handed 
neutrinos $\Psi_I({\bf1})$. 

The Majorana mass matrix $M_\nu$ of (left-handed) neutrino is induced 
from these masses $M_\R$ and $M_{\rm D}$ by 
the see-saw mechanism\cite{ref:seesaw} 
and evaluated as
\begin{eqnarray}
\bigl(M_{\nu}\bigr)_{ij}
&\sim& \bigl(M_{\rm D}^T\bigr)_{iI}
\bigl(M_\R^{-1}\bigr)_{IJ}
\bigl(M_{\rm D}\bigr)_{Jj} \nn
 &\sim&
\lambda^{\psi_i({\bf5}^*)}\left(\lambda^{\psi_I^\R}
\bigl(M_\R^{-1}\bigr)_{IJ}
\lambda^{\psi_J^\R}\right)
\lambda^{\psi_j({\bf5}^*)} 
\propto 
\lambda^{\psi_i({\bf5}^*)+\psi_j({\bf5}^*)}
\end{eqnarray}
Note here that the dependence on the 
$U(1)_X$ charges $\psi_I^\R$ of the right-handed 
neutrinos has completely dropped off.\footnote{We should however take it
account that this occurs only for a generic case and may be broken in 
particular cases in which $\bigl(M_{\rm D}^T\bigr)_{iI}$ brings about 
correlation between the left-handed neutrino index $i$ and right-handed 
one $I$.\cite{ref:BK}} Plaguing the values (\ref{eq:value5}) for $\psi 
_i({\bf5}^*)$, we thus have
\begin{equation}
M_\nu \  \propto \  
\ppmatrix{
\lambda^2 & \lambda & \lambda \\
\lambda & 1 & 1 \\
\lambda & 1 & 1 }.
\label{eq:Ours}
\end{equation}
This neutrino mass matrix happens to take the same form as one of the 
models that have been proposed by Ling and Ramond.\cite{ref:Ramond} 
This form is very interesting. 

First, this matrix implies the large 2-3 mixing in the 
diagonalization matrix $U_\nu$. The 2-3 mixing is also large in the 
charged lepton mixing matrix $U_l$ as we have seen above, and so is it 
generally in the MNS matrix $U_{\rm MNS}=U_lU_\nu^\dagger$ unless a 
cancellation occurs between $U_l$ and $U_\nu$. 

Second, it is natural to assume that three neutrino masses are not so 
degenerate accidentally. Then, the mass squared difference ratio 
(\ref{eq:dmRatio}) of solar and atmospheric neutrinos 
implies the mass ratio of the second and third neutrinos:
\begin{equation}
{m_{\nu2}\over m_{\nu3}}\ \sim\ \lambda.
\end{equation}
In order for the $M_\nu$ to reproduce this mass ratio, 
the $2\times2$ bottom-right submatrix 
of this $M_\nu$ should not be naturally-expected order 1, but 
should be $O(\lambda)$; that is, it is diagonalized by an $2\times2$ 
unitary matrix $u_\nu$ as
\begin{equation}
{\rm det}\tpmatrix{
 1 & 1 \\
 1 & 1 }\ \sim\ O(\lambda^1)\qquad \rightarrow\qquad 
u^*_\nu\tpmatrix{
 1 & 1 \\
 1 & 1 } u^\dagger_\nu\ \sim\ 
\tpmatrix{  \lambda & 0 \\
 0 & 1 }.
\end{equation}
If this is the case, 
the mass matrix $M_\nu$ takes the following form after the 
diagonalization of this $2\times2$ bottom-right submatrix:
\begin{equation}
M_\nu\qquad  \rightarrow \qquad   
\tpmatrix{
1 & 0 \\ 
0 & u_\nu^* }
M_\nu 
\tpmatrix{
1 & 0 \\ 
0 & u_\nu^\dagger} 
\ \sim \ 
\ppmatrix{
\lambda^2 & \lambda & \lambda \\
\lambda & \lambda & 0 \\
\lambda & 0 & 1 }.
\end{equation}
If we note the $2\times2$ top-left submatrix of this matrix
\begin{equation}
\tpmatrix{
\lambda^2 & \lambda \\
\lambda & \lambda },
\end{equation}
we see that this also gives the large mixing in the 1-2 sector so that 
it explains the {\em bi-large mixing}. 

Therefore, the experimental fact 
\begin{equation}
{\triangle m_\odot^2\over\triangle m_{\rm atm}^2} \ \sim\ \lambda^{2 \hbox{-} 3}
\qquad \Leftrightarrow \qquad 
{m_{\nu2}\over m_{\nu3}}\ \sim\ \lambda 
\end{equation}
necessarily implies the {\rm bi-large mixing}!

We note that a very similar neutrino mass matrix $M_\nu$ to ours
(\ref{eq:Ours}) was also proposed by Maekawa:\cite{ref:maekawa}
\begin{equation}
M_\nu \  \propto \  
\ppmatrix{
\lambda^2 & \lambda^{1.5} & \lambda^1 \\
\lambda^{1.5} & \lambda^1 & \lambda^{0.5} \\
\lambda^1 & \lambda^{0.5} & 1 }.
\label{eq:Maekawa}
\end{equation}

\section{Prediction on $U_{e3}$}

We should note that there is one more prediction in our framework, 
that is, the magnitude of the $U_{e3}\equiv(U_{\rm MNS})_{13}$:
\begin{equation}
U_{e3} \ \sim\ O(\lambda^1) \quad  \sim \quad  
\bigl(\ \underbrace{0.5}_{\lambda^{0.5}} -
 \underbrace{0.1}_{\lambda^{1.5}}\ \bigr)
\end{equation}
This is seen as follows.
First, we have
\begin{equation}
(U_l)_{11} \ \sim\ O(1), \qquad 
(U_l)_{12} \ \hbox{and} \ (U_l)_{13} \ \sim\ 
\lambda^{\psi_1({\bf5}^*)-\psi_{{2\atop3}}({\bf5}^*)} = \lambda^1,
\end{equation}
which have resulted from down-type quark masses and an 
$SU(5)$ relation.
Second, we have for 
the matrix elements of $U_\nu$,
\begin{equation}
(U_\nu)_{31} \ \sim\ \lambda^{\psi_1({\bf5}^*)-\psi_3({\bf5}^*)} = \lambda^1, \qquad 
(U_\nu)_{32} \ \hbox{and} \ (U_\nu)_{33} \ \sim\ O(1).
\end{equation}
These clearly give rise to 
$U_{e3}\equiv(U_{\rm MNS})_{13}=(U_lU_\nu^\dagger)_{13}\sim O(\lambda)$. 

This prediction gives a crucial test for the idea of Froggatt-Nielsen 
mechanism.

\section{Conclusion}

I have shown the following points in this paper:
\begin{enumerate}
\item
If we assume Froggatt-Nielsen's factorized form for the quark/lepton 
mass matrices and the $SU(5)$ structure for the $U(1)_X$ charges, an 
input of up- and down-type quark masses necessarily implies that 
the 2-3 mixing is large in the MNS matrix $U_{\rm MNS}$.
\item
If we further add the data $\sqrt{\triangle m_\odot^2/\triangle m_{\rm atm}^2} \ \sim\ \lambda$,
then, it implies that the 1-2 mixing in $U_{\rm MNS}$ is also large, 
so leading to bi-large mixing.

\item
The measurement of $U_{e3}$ will {\em confirm} or {\em kill} the 
basic idea of {\em Froggatt-Nielsen mechanism} for explaining the 
hierarchical mass structures of quarks and leptons. 

\end{enumerate}

\subsection*{Acknowledgments}

I would like to thank Masako Bando for collaborations in the work 
which this talk is based on. I also thank N.~Maekawa 
for valuable discussions. I was also inspired by stimulating 
discussions during the Summer 
Institute 2001 and 2002 held at Fuji-Yoshida.

\end{document}